\documentclass[sigconf,review=false]{acmart}

\AtBeginDocument{%
  \providecommand\BibTeX{{%
    \normalfont B\kern-0.5em{\scshape i\kern-0.25em b}\kern-0.8em\TeX}}}

\acmConference[ExHET'23]{The 2nd International Workshop on Extreme Heterogeneity Solutions}{February 25th, 2023}{Montreal, Canada}
\copyrightyear{2023}
\acmYear{2023}
\setcopyright{usgovmixed}\acmConference[ExHet'23]{The 2nd International Workshop on Extreme Heterogeneity Solutions}{February 25th, 2023}{Montreal, Canada}
\acmBooktitle{Venue, TBD}
\acmPrice{15.00}
\acmDOI{xx.xxxx/x.y}
\acmISBN{xxx-x-xxxx-xxxx-dd/yy/mm}

\usepackage{xcolor}

\usepackage{newtxmath}
\usepackage{algorithmic}
\usepackage{graphicx}
\usepackage{import}

\usepackage{multirow}
\usepackage{footnote}
\usepackage{hyperref}
\usepackage{dirtytalk}
\usepackage{caption, subcaption}
\usepackage[flushleft]{threeparttable}
\colorlet{myOrange}{green!10!orange!90!}

\usepackage{tikz}
\newcommand*\circled[1]{\tikz[baseline=(char.base)]{
            \node[shape=circle,draw,inner sep=2pt] (char) {#1};}}

\colorlet{punct}{red!60!black}
\definecolor{background}{HTML}{EEEEEE}
\definecolor{delim}{RGB}{20,105,176}
\colorlet{numb}{magenta!60!black}
\usepackage{listings}

\definecolor{zhen}{RGB}{26, 153, 62}

\definecolor{codegreen}{rgb}{0,0.6,0}
\definecolor{codegray}{rgb}{0.5,0.5,0.5}
\definecolor{codepurple}{rgb}{0.58,0,0.82}
\definecolor{backcolour}{rgb}{0.95,0.95,0.92}
\definecolor{cosmiclatte}{rgb}{1.0, 0.97, 0.91}
\usepackage[justification=centering]{caption}
\usepackage{listings}
\lstdefinestyle{mystyle}{
    backgroundcolor=\color{cosmiclatte},   
    commentstyle=\color{codegreen},
    keywordstyle=\color{magenta},
    numberstyle=\tiny\color{codegray},
    stringstyle=\color{codepurple},
    basicstyle=\ttfamily\footnotesize,
    breakatwhitespace=false,         
    breaklines=true,                 
    captionpos=t,                    
    keepspaces=true,                 
    numbers=left,                    
    numbersep=5pt,                  
    showspaces=false,                
    showstringspaces=false,
    showtabs=false,                  
    tabsize=2
}
\usepackage{multirow,tabularx,verbatim}
\newcolumntype{Y}{>{\centering\arraybackslash}X}
\renewcommand\arraystretch{2}

\newcolumntype{P}[1]{>{\centering\arraybackslash}p{#1}}

\begin{document}

\title{Transfer Learning Across Heterogeneous Features For Efficient Tensor Program Generation}


\author{Gaurav Verma}
\affiliation{%`
  \institution{Stony Brook University}
  \city{Stony Brook}
  \state{New York}
  \country{USA}
  }
\email{gaurav.verma@stonybrook.edu}
  
\author{Siddhisanket Raskar}
\affiliation{%`
  \institution{Argonne National Laboratory}
  \city{Lemont}
  \state{Illinois}
  \country{USA}
  }
  \email{sraskari@anl.gov}
  
\author{Zhen Xie}
\affiliation{%`
  \institution{Argonne National Laboratory}
  \city{Lemont}
  \state{Illinois}
  \country{USA}
  }
  \email{zhen.xie@anl.gov}
  
\author{Abid M. Malik}
\affiliation{%`
  \institution{Brookhaven National Laboratory}
  \city{Upton}
  \state{New York}
  \country{USA}
  }
\email{amalik@bnl.gov}
  
\author{Murali Emani}
\affiliation{%`
  \institution{Argonne National Laboratory}
  \city{Lemont}
  \state{Illinois}
  \country{USA}
  }
\email{memani@anl.gov}
  
\author{Barbara Chapman}
\affiliation{%`
  \institution{Stony Brook University}
  \city{Stony Brook}
  \state{New York}
  \country{USA}
}
\email{barbara.chapman@stonybrook.edu}

%%
%% By default, the full list of authors will be used on the page
%% headers. Often, this list is too long and will overlap
\renewcommand{\shortauthors}{Verma, et al.}

\begin{abstract}
Tuning tensor program generation involves searching for various possible program transformation combinations for a given program on target hardware to optimize the tensor program execution. It is already a complex process because of the massive search space, and exponential combinations of transformations make auto-tuning tensor program generation more challenging, especially when we have a heterogeneous target. In this research, we attempt to address these problems by learning the joint neural network and hardware features and transferring them to the new target hardware. We extensively study the existing state-of-the-art dataset, TenSet, perform comparative analysis on the test split strategies and propose methodologies to prune the dataset. We adopt an attention-inspired approach for tuning the tensor programs enabling them to embed neural network and hardware-specific features. Our approach could prune the dataset up to 45\% of the baseline without compromising the Pairwise Comparison Accuracy (PCA). Further, the proposed methodology can achieve on-par or improved mean inference time with 25\%-40\% of the baseline tuning time across different networks and target hardware.

\end{abstract}



\maketitle
\section{Introduction}
Deep neural networks (DNN) applications are ubiquitous across multiple artificial intelligence domains, including industrial and scientific disciplines. They form the backbone of many existing and emerging applications. The DNN development is rapidly advancing due to the 
% supported by the development of 
the capabilities of computing hardware and domain-specific accelerators that make the execution of tensor programs efficient by providing hand-tuned deep learning (DL) libraries. These
%hand-tuned 
libraries are often not scalable. 
Tensor compilers such as XLA~\cite{xla}, TVM~\cite{chen2018tvm}, and Glow~\cite{rotem2018glow}  %and TACO~\cite{kjolstad2017taco} 
facilitate fine-grained hardware-independent (high-level) and dependent (low-level) optimizations to the input computation graphs.
% To facilitate fine-grained hardware-independent (high-level) and dependent (low-level) optimizations to the input computation graphs, tensor compilers like XLA~\cite{xla}, TVM~\cite{chen2018tvm}, Glow~\cite{rotem2018glow}, and TACO~\cite{kjolstad2017taco} offer a platform. 
A tensor compiler analyzes the computation graphs and applies various optimizations at different stages. Additionally, it inputs subgraphs or mathematical expressions and selects the optimized low-level implementation generating a tensor program. Tensor program generation refers to automatically generating code for tensor programs. 
% of the same. 

The deep neural architecture has evolved manifolds from simple neural networks to convoluted ones, followed by recurrent ones to massively large models.% such as Megatron-Turing Natural Language Generation (MT-NLG)~\cite{shoeybi2019megatron} \textcolor{zhen}{we do not test large model}. 
Advancements in hardware such as GPUs and domain-specific accelerators %\textcolor{zhen}{we do not use and evaluate that} 
and DL frameworks like TensorFlow and PyTorch offer optimized kernel support facilitating DL innovations. With advances in DNN architectures and backend hardware, the search space of compiler optimizations has grown manifold. The vast search space consisting of loop optimizations like tiling, vectorization, etc., limits the use of data-driven approaches to auto-tune the tensor compilers and generate efficient tensor programs. With automatic tuning of tensor compilers to generate performant kernel plans becoming prevalent, proposed methodologies based on data-driven cost modeling and intelligent techniques require significant amounts of hardware data to learn cost models. Often, these datasets are specific to the nature of the experiments in the HPC domain. These resource-intensive techniques are hindered by the large number of kernels required upon introducing new hardware. In such a scenario, the cost model or a tuner requires retraining from scratch.  

Additionally, almost all of these cost models~\cite{chen2018learning, ryu2022one} % \ME{any citations?} 
are based on the train and test data drawn from identical probability distributions, while the source and target compute hardware are the same. However, with the advancement in heterogeneous hardware systems, such as different generations of CPUs and GPUs, it may not be practical to have this assumption. For example, a tuner trained for a DL workload on a CPU may not generate efficient tensor programs on GPUs. Transfer learning has been advantageous in mitigating these limitations. It can learn the context from the neural network tasks and apply them to the new context. 

Hence, the need for a transfer learning-based methodology requiring lesser data and quick adaptability to the new hardware is paramount. The cost models trained on limited hardware or neural network tasks lack transfer learning capabilities. Consequently, it is efficient to map the heterogeneous feature space across devices and fine-tune only the required features applicable instead of retraining from scratch. Where hand-tuned libraries fall short of providing optimized support to new hardware and operators, auto tuners requiring lesser data to learn can reduce the tuning time and the time required for online device measurement by focusing on learning high-performance kernels. Recent works have proposed transfer learning methods for the same source and target hardware~\cite{zheng2021tenset,ryu2022one,gibson2022transfer}. We suggest heterogeneous transfer learning by mapping the kernel as a feature space in a new context. 

This paper analyzes the current methods used to generate tensor programs using transfer learning for CPU and GPU-based systems. We conduct probabilistic and exploratory analyses to achieve comparable results using less data than the baseline across various split strategies. We propose a transfer learning approach to generate efficient tensor programs with less tuning time and fewer kernel measurements across heterogeneous hardware. The significant contributions of this paper are as follows:
\begin{itemize}
 \item perform extensive study on the existing work to extract and learn from the joint significant neural network and hardware features
    \item define the proposed methodology based on fewer kernels measurements and efficient transfer learning
    \item implement an optimized tuner based on the key points above and present results for heterogeneous transfer learning. We could achieve on-par or better mean inference time with 3x-5x improved tuning time and up to 47\% reduced dataset.
\end{itemize}

The remainder of the paper is organized as follows: Section 2 gives the requisite background to understand the problem and presents the related work in this area. Section 3 describes the proposed methodology used in the study. Sections 4 and 5 discuss experimentation and results, respectively. Section 6 provides a conclusion and potential future steps.

\section{BACKGROUND AND RELATED WORK}
\subsection{Cross-device Learning}
The search space in cross-device learning is enormous, ranging from the orders of millions (CPU) to billions (GPU). It leads to ample search space and incurs high costs in terms of search time. Transferring auto-schedule knowledge from pre-tuned kernels to untuned kernels can play a significant role. As discussed here, researchers have proposed solutions to address various aspects of transfer learning~\cite{gibson2022transfer}. In~\cite{gibson2022transfer, verma2022towards}, authors have classified tasks into classes to have a better selection of optimizations. ~\cite{mendis2019ithemal} uses a hierarchical LSTM–based approach to predict throughput based on the opcodes and operands of instructions in a basic block. The authors propose a solution that is easily portable across various processor micro-architectures. Other research uses a cost model query optimizer to improve resource utilization and lower operational costs ~\cite{siddiqui2020cost}. In ~\cite{zhang2020autosync}, the authors propose an end-to-end pipeline to optimize synchronization strategies given model structures and resource specifications, lowering the bar for data-parallel distributed ML across devices. 
%~\cite{zhao2022moses} proposes a solution based on the lottery ticket hypothesis to select transferable features across devices via domain adaptation. 
Another work ~\cite{ahn2020chameleon} employs reinforcement learning to develop an adaptive sampling algorithm to alleviate costly hardware measurement. Our work proposes efficient pruning of datasets to learn joint kernel and hardware features. It enables efficient tuning by considering prominent kernels. 

\subsection{Machine Learning-based Auto Tuners}
ML-based approach to automatically optimize tensor programs is a heavily researched area ~\cite{chen2018learning, kaufman2021learned, cummins2017end} 
%%adams2019learning, baghdadi2021deep, jung2021deepcuts
focusing on tensor compilers % ~\cite{rotem2018glow, chen2018tvm, xla} 
and DL workloads~\cite{verma2021performance}.
~\cite{zheng2020ansor} is a framework to generate tensor programs for the DL workloads. It explores optimization combinations by sampling programs from a hierarchical search space representation and optimizes multiple sub-graphs using a task scheduler. 
%Another research that explores schedules on heterogeneous systems is ~\cite{zheng2020flextensor}. They propose statistical analysis and ml-based heuristics to achieve it.
~\cite{steiner2021value} employs LSTM to formulate the scheduling process as a sequence of optimization choices. 
%Similarly, ~\cite{ashouri2016cobayn} proposes a solution using the bayesian network. 
Furthermore, ~\cite{whaley1998automatically} described an approach for automatically generating and optimizing numerical software for processors with deep memory hierarchies and pipelined functional units. Such an approach makes its adoption across server and mobile platforms feasible.

Additionally, authors have employed reinforcement learning (RL) in various works. % ~\cite{cummins2022compilergym}. %zhu2022roller, ansel2014opentuner, nakandala2020tensor
%In ~\cite{zhang2020dynatune}, authors consider a Multi-Armed Bandit (MAB) model for the tensor program optimization problem. 
In ~\cite{mendis2019compiler}, the authors explore the feasibility of formulating the integer linear programming solver into a graph neural network-based policy to auto-generate a vectorization scheme. 
Further, domain-specific compilers like COMET~\cite{mutlu2022comet}, JAX~\cite{jax2018github}, and NWChem~\cite{valiev2010nwchem} are actively researched to optimize the underlying program's execution. ~\cite{mutlu2022comet} performs domain-specific, hardware-agnostic optimizations that rely on high-level semantic information and are applied at high-level IRs. 
%~\cite{huang2019autophase} proposes an RL framework to address the phase ordering problem for high-level synthesis programs. 
Lately, in ~\cite{ryu2022one}, authors have proposed a one-shot tuner for the tensor compilers. Its limitation is that it considers only task-specific information, which prevents it from being applied for transfer learning. On the other hand, we employ neural network and hardware platform information so that the auto-tuner can learn better. 

\section{Design And Implementation}
\begin{figure}[ht!] %!t
\centering
\includegraphics[width=1\linewidth]{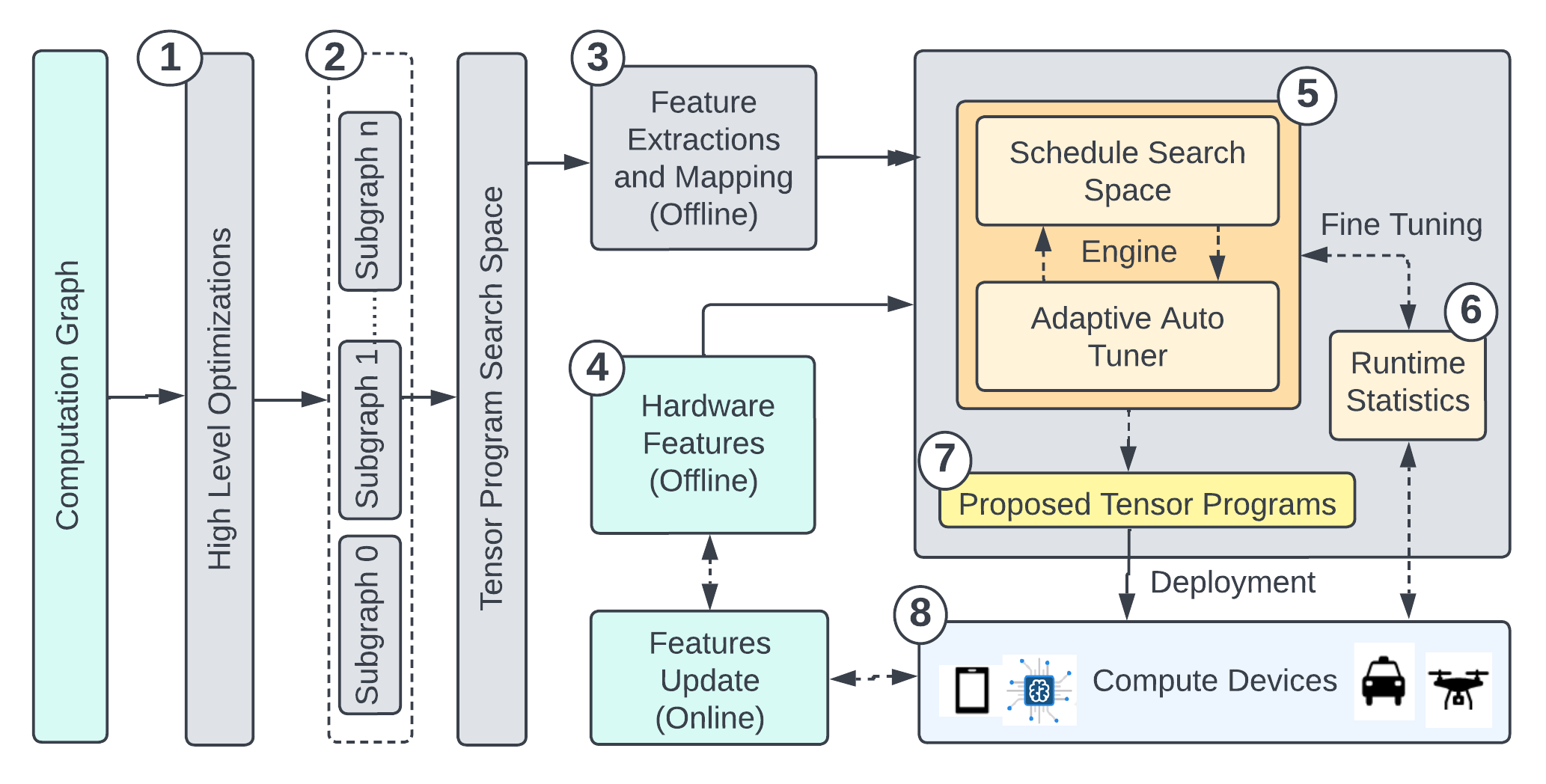}
\vspace{-0.3cm}
\caption{Overview of the Proposed Framework}
\label{ref:overview}
\end{figure}

This section unfolds our methodology as follows: Section~\ref{sec:sysOver} provides an overview of the entire flow and individual components, Section~\ref{sec:featureSampling} elaborates on the proposed optimizations and opportunities for efficient heterogeneous transfer learning with less dataset, and Section~\ref{sec:opt} explains the adaptive auto-tuner architecture. 

\subsection{An Overview}%\ME{rename `system'}}
\label{sec:sysOver}
Figure~\ref{ref:overview} provides an end-to-end flow of our framework. We  used TVM \texttt{v0.8dev0} as a base to present the heterogeneous transfer learning. The user leverages the TVM API to \circled{1} provide the computation framework in the supported format (TensorFlow, PyTorch, etc.). It further \circled{2} undergoes high-level optimization and subgraph partitioning to generate subgraphs of a smaller order. The induced subgraphs form the search space for the feature extractions. Additionally, as a feature set, \circled{3} domain-specific information is extracted from these subgraphs, like kernel dimensions, tensor operations, etc. For each data-point entry or kernel, \circled{4} we store the hardware information like hardware architecture, maximum number of threads, registers and threads per block, and shared memory. This forms part of the static hardware dataset. We perform a probabilistic and exploratory study on this feature set to identify the features of high importance. The hardware characterization assists in mapping the features from source to target in case of a different hardware than the source. We \circled{5} train auto-tuner on this dataset by extending a one-shot tuner~\cite{ryu2022one}. Once the auto-tuner is trained, it can be used to perform auto-generation of tensor programs on the target device with or without re-training. If the user wants to fine-tune the auto tuner, we provide a capability to \circled{6} fine-tune the auto tuner using online hardware features. We have experimented with selective tasks retraining, and are still working on a methodology for selective feature training. %, like Lottery Ticket Hypothesis-based technique~\cite{zhao2022moses}. 
The non-parametric approach reduces the retraining time and size of the dataset required. Also, we used attention heads to support memory-augmented fine-tuning using bidirectional LSTM. Since the training strategy is based on the one-shot paradigm, the baseline auto-tuner model undergoes a one-time complete training phase alleviating the large data requirement on the target device for full retraining. Hence \circled{7} proposed tensor programs are \circled{8} deployed on the target hardware and their performance is evaluated. 

We used the TenSet dataset~\cite{zheng2021tenset} for the baseline measurements. As per authors claims, the dataset can be leveraged for  generating the tensor programs using transfer learning. The basis is diversity in the dataset leading to generalization and multi-platform performance data points (measured on CPUs of Intel, AMD and ARM and NVIDIA GPUs). Section ~\ref{sec:opt} discusses the applicability of these bases for efficient transfer learning using the proposed auto-tuner. For more information on the dataset, one can refer to ~\cite{zheng2021tenset}. Table ~\ref{tab:hardwarePlatform} summarizes the hardware and associated characteristics that we have considered in our study. 

\begin{table}[ht!]
    \caption{Compute Hardware Description}
    \vspace{-0.2cm}
    \label{tab:hardwarePlatform}
\resizebox{\columnwidth}{!}{
\renewcommand{\arraystretch}{1.1}
    \begin{tabular}{|l|c|c|c|} \hline 
    \textbf{Hardware Platform} & \textbf{Processor} & \textbf{Remarks} \\
    \hline \hline
    Intel Platinum 8272CL @ 2.60GHz & CPU & 16 cores, AVX-512 \\
    \hline
    AMD EPYC 7452 @ 2.35GHz & CPU & 4 cores, AVX-2\\
    \hline
    ARM Graviton2 & CPU & 16 cores, Neon\\
    \hline
    NVIDIA Tesla T4 & GPU & Turing Architecture\\
    \hline
    NVIDIA GeForce RTX 2080 & GPU & Turing Architecture\\
    \hline
    NVIDIA A100 & GPU & Ampere Architecture\\
    \hline
    NVIDIA A40 & GPU & Ampere Architecture\\
    \hline
    Intel Gold 5115 @ 2.40GHz & CPU & 40 cores, Xeon \\
    \hline
    \end{tabular}}
\end{table}

\subsection{Hardware-Aware Kernel Sampling}
\label{sec:featureSampling}

We have extensively studied and conducted an experimental examination of the TenSet %~\cite{zheng2021tenset} 
dataset to understand the neural network and hardware features that can influence the metrics like flops, throughput, and latency. The dataset consists of over 13,000 tasks from 120 networks measured on six hardware platforms for the metrics like throughput and latency under different neural network and hardware parameters resulting in over 52 million measurements. As shown in Table~\ref{tab:hardwarePlatform}, we have considered the first four platforms from the dataset to learn the task, schedules applied to them, and associated performance along with the hardware parameters. The latter half of the hardware is used to evaluate and establish the usefulness of cross-device or transfer learning. In addition to the neural network information like tensor operation and input and output shape, we have also considered hardware parameters, as shown in Table~\ref{tab:hardwareParam}. Here, we have presented hardware features only for CPU and GPU but it can be extended to other heterogeneous devices. Analyzing the dataset, we identified that most high-performant kernels are associated with specific hardware parameters. Such probabilistic sampling also mitigates skewed kernel selection, avoiding kernels that can lead to lower FLOPs or invalid computation graphs. We removed the measurements and kernels which were invalid or low-performing. In the existing cost modeling, it is observed that the large search space selected via random sampling of kernels causes performance regression. 

\begin{table}[ht!]
    \caption{Hardware Parameters Considered While Training}
    \vspace{-0.2cm}
    \label{tab:hardwareParam}
    \resizebox{\columnwidth}{!}{%
\renewcommand{\arraystretch}{1.7}
    \begin{tabular}{|l|c|c|c|c|} \hline 
\textbf{Hardware Parameter} & \textbf{Definition}	& \textbf{Hardware Class}	& \textbf{Value (bytes)} \\
    \hline \hline
    cache\_line\_bytes & chunks of memory handled by the cache & CPU; GPU	& 64\\
    \hline
    max\_local\_memory\_per\_block	& maximum local memory per block in bytes &	GPU	& 2147483647\\
    \hline
    max\_shared\_memory\_per\_block	& maximum shared memory per block in bytes	& GPU & 49152\\
    \hline
    max\_threads\_per\_block & maximum number of threads per block	& GPU	& 1024\\
    \hline
    max\_vthread\_extent &	maximum extent of virtual threading	& GPU &	8\\
    \hline
    num\_cores	& number of cores in the compute hardware	& CPU	& 24\\
    \hline
    vector\_unit\_bytes	& width of vector units in bytes	& CPU; GPU	& 64; 16\\
    \hline
    warp\_size	& thread numbers of a warp	& GPU	& 32\\
    \hline
    \end{tabular}}
\end{table}

\begin{table*}[ht!]
    \centering
\caption{Neural-Network And Hardware Features Characterization}
\vspace{-0.2cm}
    \label{tab:hardwareCharacter}
    \resizebox{\textwidth}{!}{
    \renewcommand{\arraystretch}{1}
    \begin{tabular}{|l|c|c|c|c|c|c|c|c|c|c|}
    \hline
    \multirow{2}{*}{Sampled Kernels} & \multicolumn{2}{|c|}{\#Kernel\_Shapes} &
    \multicolumn{2}{|c|}{Max GFLOPs} &
     \multirow{2}{*}{Tensor Shape} & 
     \multicolumn{4}{|c|}{Mean Execution Time (ms)} \\
    \cline{2-5}\cline{7-10}
         & CPU & GPU  & CPU & GPU &  & EPYC-7452	& Graviton2	& Platinum-8272	& T4 \\
    \hline\hline
    T\_add  &  229  &  388  &  8.59  &  8.59  &  [4, 256, 1024]  &  180.97  &  81.25  &  92.86  &  4.31 \\
	\hline
    Conv2dOutput  &  60  &  27  &  1.20  &  1.07  &  [4, 64, 64, 32]  &  40.94  &  14.21  &  19.11  & 2.07 \\
	\hline
    T\_divide  &  24  &  69  &  0.003  &  0.003  &  [8, 1, 1, 960]  &  0.07  &  0.05  &  0.11  &  0.10 \\
	\hline
    T\_fast\_tanh  &  9  &  9  &  0.008  &  0.008  &  [4, 1024]  &  0.43  &  0.43  &  0.53  &  0.97 \\
	\hline
    T\_multiply  &  105  &  150  &  8.92 &  8.92  &  [4, 256, 4096]  &  320.74  &  48.08  &  95.65  & 0.55 \\
	\hline
    T\_relu  &  300  &  1257  &  73.46  &  73.46  &  [4, 144, 72, 8, 64]  &  0.52  &  5.70  &  0.72  &  0.23 \\
	\hline
    T\_softmax\_norm  &  27  &  27  &  0.016  &  0.016  &  [4, 16, 256, 256]  &  1.01  &  2.78  &  4.08  & 0.19 \\
	\hline
    T\_tanh  &  9  &  9  &  0.905  & 0.629  &  [8, 96, 96, 3]  &  5.55  &  33.48  &  50.55  &  0.16 \\
	\hline
    conv2d\_winograd  &  0  &  33  &  NA  &  0.868  &  NA  &  NA  & NA  &  NA & 0.93 \\	
    \hline
    \end{tabular}}
    \begin{tablenotes}
    \item[*] \footnotesize{*CPU: EPYC-7452, Graviton2, Platinum-8272}; \footnotesize{*GPU: T4}; \footnotesize{*NA: operator is not present in the considered CPU dataset}
    \footnotesize{**Measured on Nvidia GeForce RTX 2080}
    \end{tablenotes}
\end{table*}

We pruned the vast search space for the tensor program generation. The search task based on random sampled kernels' measurements is unreliable when the search space is not rich. We observed that specific hardware parameters, like the number of cores, are less valuable features than flop count regarding latency. It is mainly because of the need for diversified kernel combinations and hardware features in the considered dataset. Hence, we evaluated the combination of FLOPs count, kernel shape, and execution time on a given hardware for various tensor operations. As shown in Table~\ref{tab:hardwareCharacter}, we observed that the CPU and GPU behave similarly when selecting the best shape for a kernel. Although, the execution time differs significantly based on the compute hardware. Hence, we sampled the kernels based on joint exploration of tensor operations, kernel shapes, and hardware parameters based on the FLOPs count and execution time. The extracted initial kernels are from six diversified neural networks to encompass prominent classes. Still, it will be an ever-growing list, and we are working on an intelligent algorithm to achieve the same whenever introducing a new network class. Experimenting with \texttt{rmse} and \texttt{ranking loss}, we established that \texttt{rmse} performs better based on inference time. Hence, we chose \texttt{rmse} for our tuner's performance evaluation. We learned from the measurement records' costs of the selected kernels that it contains local minima. To avoid local minima, we leveraged simulated annealing implemented in TVM. But it is computationally heavy and takes significant time to find global optima. We aim to optimize it in our future work. % steps. 

\subsection{Auto-tuner Architecture}
\label{sec:opt}
We have extended the one-shot tuner to experiment with a joint neural network and hardware features as part of the task. Our tuner consists of bidirectional LSTM and attention head to learn from the sequential data. After testing multiple configurations, we chose the near-optimal values based on empirical evaluation, including a batch size of 16, 200 epochs, and other hyperparameters listed in Table~\ref{tab:hyperparam}.
\begin{table}[ht!]
    \caption{Auto-Tuner's Hyperparameters}
    \vspace{-0.2cm}
    \label{tab:hyperparam}
\resizebox{\columnwidth}{!}{
\renewcommand{\arraystretch}{0.7}
    \small
    \begin{tabular}{|l|c|c|} \hline 
    \textbf{Hyperparameter} & \textbf{Value} \\
    \hline \hline
    Batch & 1, 4, {\bf 16}, 64, 256 \\
    \hline
    Epoch & 100, {\bf 200}, 400 \\
    \hline
    Learning Rate & 1e-3 \\
    \hline
    \#Bidirectional LSTM Layers & 3 \\
    \hline
    Attention Head (fine tuning) & 2 \\
    \hline
    \#Unrolling Steps for Attention Head & 2 \\
    \hline
    Optimizer & Adam \\
    \hline
    \end{tabular}}
\end{table}

\section{Evaluation}
\subsection{Experimental Setup}

{\bf Platform:} We used heterogeneous architectures like Nvidia GPUs- RTX 2080, A100, A40, and Intel Xeon CPU for our study experiment. We chose different generations of NVIDIA GPUs to study the impact of architectural differences. \\
{\bf Dataset and Model:} This work use TVM v0.8dev0 and PyTorch for implementations. We used XGBoost (XGB), multi-layer perception (MLP), and LightGBM (LGBM) based tuners as a baseline.\\
{\bf Baseline:} For the baseline, we used the TenSet dataset, commit 35774ed. Based on the previous  work~\cite{zheng2021tenset}, we have considered 800 tasks with 400 measurements as the baseline. We used Platinum-8272 for the CPU dataset and Nvidia Tesla T4 for the GPU dataset.%The CPU dataset measurements were obtained from Platinum-8272 and the GPU dataset measurements were sourced from Nvidia Tesla T4.

\subsection{Dataset Sampling}

\begin{table*}[]
\caption{Reduction In Dataset Size And Train-time (by split strategies)}
\vspace{-0.2cm}
\label{tab:sizetimereduction}
\renewcommand{\arraystretch}{1.2}
\resizebox{\textwidth}{!}{
\begin{tabular}{|c|c|c|ccc|ccc|ccc|}
\hline
\multirow{2}{*}{Target Hardware} & \multirow{2}{*}{Dataset} & \multirow{2}{*}{Size} & \multicolumn{3}{c|}{XGBoost (train-time (sec))}                                            & \multicolumn{3}{c|}{MLP (train-time (sec))}                                             & \multicolumn{3}{c|}{LightGBM (train-time (sec))}                                            \\ \cline{4-12} 
                                 &                          &                       & \multicolumn{1}{c|}{within\_task} & \multicolumn{1}{c|}{by\_task} & by\_target & \multicolumn{1}{c|}{within\_task} & \multicolumn{1}{c|}{by\_task} & by\_target  & \multicolumn{1}{c|}{within\_task} & \multicolumn{1}{c|}{by\_task} & by\_target  \\ \hline\hline
\multirow{2}{*}{GPU}             & Baseline                 & 16G                   & \multicolumn{1}{c|}{1504}       & \multicolumn{1}{c|}{1440}   & 454 & \multicolumn{1}{c|}{3000}       & \multicolumn{1}{c|}{2434}   & 3150 & \multicolumn{1}{c|}{1574}       & \multicolumn{1}{c|}{780}    & 4680 \\ \cline{2-12} 
                                 & Sampled                  & 9G                    & \multicolumn{1}{c|}{1406}       & \multicolumn{1}{c|}{1169}   & 339 & \multicolumn{1}{c|}{1968}       & \multicolumn{1}{c|}{1655}   & 2464 & \multicolumn{1}{c|}{1175}       & \multicolumn{1}{c|}{595}    & 3637 \\ \hline
\multirow{2}{*}{CPU}             & Baseline                 & 11G                   & \multicolumn{1}{c|}{1490}       & \multicolumn{1}{c|}{1265}   & 428 & \multicolumn{1}{c|}{3143}       & \multicolumn{1}{c|}{2623}   & 2043 & \multicolumn{1}{c|}{1131}       & \multicolumn{1}{c|}{636}    & 3946 \\ \cline{2-12} 
                                 & Sampled                  & 6.8G                  & \multicolumn{1}{c|}{905}        & \multicolumn{1}{c|}{780}    & 354 & \multicolumn{1}{c|}{2091}       & \multicolumn{1}{c|}{1672}   & 1270 & \multicolumn{1}{c|}{489}        & \multicolumn{1}{c|}{387}    & 2435 \\ \hline
\end{tabular}}
\end{table*}

As discussed in Section~\ref{sec:featureSampling}, we have sampled the dataset. By employing the data sampling strategies based on the feature's importance in terms of FLOPs count, we could reduce the GPU dataset by 43\% and the CPU dataset by 47\%. As shown in Table~\ref{tab:sizetimereduction}, we could gain the training time overall. 

While sampling the dataset, it is essential to ensure the accuracy of the resultant cost model or tuner trained over the sampled dataset. Hence, we compared the top-1 and top-5 accuracy with the pairwise comparison accuracy (PCA). To explain briefly, if $y$ and $\hat{y}$ are actual and predicted labels, then we calculate the number of correct pairs, $CP$, by performing elementwise $xor$, followed by elementwise $not$ on $y$ and $\hat{y}$. Further, we take the sum of the upper triangular matrix of the resultant matrix. The PCA is calculated then using equation~\ref{eqn:1}.
\begin{equation}
\begin{aligned}
PCA = CP / (n*(n-1)/2); n = len(\hat{y})
\end{aligned}
\label{eqn:1}
\end{equation} 
The cost models trained on the baseline and sampled dataset performed on par. 

To have a fair comparison 
%n apple-to-apple comparison, 
we have trained XGB, MLP, and LGBM tuners on the baseline and sampled data using three split strategies as explained here: 
\begin{itemize}
    \item \emph{within\_task}: the dataset is partitioned into train and test based on the measurement record. Once the features are extracted for each task, it is shuffled and randomly partitioned.  
    \item \emph{by\_task}: a learning task is used to partition the dataset randomly based on the features of the learning task.
    \item \emph{by\_target}: partitioning is performed based on the hardware parameters
\end{itemize}
To avoid skewed sampling, tasks with too few measurements were excluded. Further, we have considered the tasks based on the FLOPs of tensor operations occurrence probability as shown in  %Table~\ref{tab:hardwareCharacter} 
Table\ref{tab:hardwareCharacter}. The latency and throughput of these tasks are recorded by executing them on the computing hardware. Table~\ref{tab:sizetimereduction} presents the gain in time-to-train for the sampled dataset. It can be seen that in the case of CPUs, the gain is up to 56\%  for LGBM while using \emph{within\_task} split strategy during training. GPUs also have shown an increase of up to 32\%.

\subsection{Tensor Program Tuning}
% \todo{Sid: can you write this section? It need not be long}
%- discuss tuning results on different HW
%- existing methodologies results
%- batch size variation, maybe!  
Here, we discuss metrics to show the effectiveness of our proposed approach over the baseline. 

\begin{table}[ht!]
\caption{Pairwise Comparison Accuracy}
\vspace{-0.2cm}
\label{tab:pca}
\resizebox{\columnwidth}{!}{
\renewcommand{\arraystretch}{1.1}
    \begin{tabular}{|l|c|c|c|c|} \hline 
    \multirow{2}{*}{Hardware} & \multirow{2}{*}{Dataset} & \multicolumn{3}{c|}{Split Scheme} \\
    \cline{3-5}
    &  & By Target & Within Task & By Task \\
    \hline\hline
    \multirow{2}{*}{NVIDIA A100 GPU}  & Baseline & 0.8476 & 0.8931 & 0.8565 \\ 
    \cline{2-5}
    & Sampled & 0.844 & 0.8953 & 0.8534 \\
    \hline 
    \multirow{2}{*}{Intel Xeon CPU} & Baseline & 0.8477 & 0.8506 & 0.8651 \\ 
    \cline{2-5}
    & Sampled & 0.8434 & 0.8456 & 0.861 \\
    \hline
    \end{tabular}}
    
\end{table}

Table~\ref{tab:pca} shows the Pairwise Comparison Accuracy(PCA) earlier discussed in Equation~\ref{eqn:1} for each split scheme for our sampled dataset with baseline dataset for NVIDIA A100 GPU and Intel Xenon CPU. We observe that accuracy hardly changes with the sampled dataset. We also observe a similar trend for other architectures used in this study. %Please refer to our GitHub\footnote{\url{https://github.com/xintin/TransferLearn\_HetFeat\_TenProgGen}} %~\cite{github_repo}  

\begin{table}[ht!]
\caption{Inference Time Comparison (Seconds)}
\vspace{-0.2cm}
\label{tab:inf_time}
\resizebox{\columnwidth}{!}{
\renewcommand{\arraystretch}{1.5}
    \begin{tabular}{|l|c|c|c|c|} \hline 
    \multirow{2}{*}{Target Hardware} & \multicolumn{2}{c|}{Baseline Dataset} & \multicolumn{2}{c|}{Sampled Dataset} \\
    \cline{2-5}
    &  W/o Transfer Tuning & W/ Transfer Tuning & W/o Transfer Tuning & W/ Transfer Tuning \\
    \hline\hline
    A100 GPU  & 578	& 391	& 585	& 400 \\ 
    \hline
    A40 GPU  & 627 & 416	& 599	& 175 \\ 
    \hline
    RTX2020 GPU  & 18.67 & 27.68 & 17.37	& 841.74 \\ 
    \hline
    Xeon CPU  & 91.34	& 282.2	& 85.22 & 189.25 \\ 
    \hline

    \end{tabular}}
    
\end{table}

Table~\ref{tab:inf_time} shows the inference times for baseline and sampled datasets with and without transfer tuning for NVIDIA A100, A40, RTX 2020 GPUs, and Intel Xeon CPU. We observe significant advantages with the sampled dataset as the inference times are much lower than the baseline dataset. The numbers discussed here are for the XGBoost tuner. Please refer to our GitHub\footnote{\url{https://github.com/xintin/TransferLearn\_HetFeat\_TenProgGen}} %~\cite{github_repo} 
for inference numbers using multi-layer perception (MLP) and LightGBM (LGBM) based tuners as well inference numbers for various batch sizes (1,2,4,8) and detailed logs for different architectures used in this study. We observe similar trends advantageous to the sampled dataset. 

\subsection{Evaluation of Heterogeneous Transfer Learning}
Different transformations can be applied for a given compute graph consisting of tensor operation and input and output tensor shapes, varying their performance on target hardware. For example, in a conv2D tensor operation, tiling is dictated by whether the target is GPU or CPU because of grid and block size bound in GPUs. A large tiling size that may be valid in CPU may be invalid in GPU. Also, not all combinations are performant. We learned the joint-optimized schedules for a kernel and hardware using the TVM auto-scheduler. Then, we applied it to similar untuned kernels using the attention mechanism. To make it efficient, we sort the kernels as per their occurrences and total contribution to the FLOPs count. Then we tune the selected few significant tensor operations. 

We have evaluated our methodology using three architecturally different networks on CPU and GPU. Contrasting to the baseline, where the tasks are fetched randomly tuned, we select the tasks contributing more to the FLOPs. As shown in Table~\ref{tab:ourtuner}, we could achieve the on-par mean inference time with significantly reduced tuning time. On CPU, for ResNet\_50, we could achieve 30\%, MobileNet\_50 70\%, and Inception\_v3 90\% reduction in time on CPU. ResNet\_50 suffered performance regression due to a lack of matching kernel shapes in the trained dataset for the given hardware. Whereas on GPU, we could achieve 80\%-90\% across all the networks. Here, we used the trainer tuned on features from neural networks and hardware.

\begin{table}[ht!]
\caption{Evaluation Of Proposed Tuner}
\vspace{-0.2cm}
\label{tab:ourtuner}
\resizebox{\columnwidth}{!}{
\renewcommand{\arraystretch}{1.5}
    \begin{tabular}{|c|c|c|c|c|c|} \hline 
    \multirow{2}{*}{Target Hardware} & \multirow{2}{*}{Network}  & \multicolumn{2}{c|}{W/o Transfer Tuning }                                            & \multicolumn{2}{c|}{W/ Transfer Tuning }      \\ 
\cline{3-6} 
 & & Time-to-Tune & Mean Inf. Time & Time-to-Tune & Mean Inf. Time \\
 \hline\hline
\multirow{3}{*}{CPU} & ResNet\_50 & 128 & 11.93 & 86 & 12.12  \\
\cline{2-6}
& MobileNet\_v3 & 236 & 5.48 & 71 & 5.57 \\
\cline{2-6}
& Incpetion\_v3 & 614 & 75.27 & 61 & 73.80 \\
\hline
\multirow{3}{*}{GPU} & ResNet\_50 & 817	& 3.79	& 226	& 3.78  \\
\cline{2-6}
& MobileNet\_v3 &  1092 & 1.72 & 136 & 1.75 \\
\cline{2-6}
& Incpetion\_v3 & 2510	& 28.72	& 191	& 28.73 \\
\hline
\end{tabular}}
\begin{tablenotes}
    \item[*] \footnotesize{*CPU: Intel Xeon}; \footnotesize{*GPU: A100}; \footnotesize{*w/o: without; w/: with; tune time (sec); inf time (ms)}
    \end{tablenotes}
\end{table}

% \begin{figure}[ht!] %!
% %\centering
% \includegraphics[width=1\linewidth]{assets/train_convergence.png}
% \caption{Comparing Training Convergence of Tuners}
% \label{ref:convergence}
% \footnotesize{*experimented on Nvidia RTX 2080}
% \end{figure}

\begin{figure*}
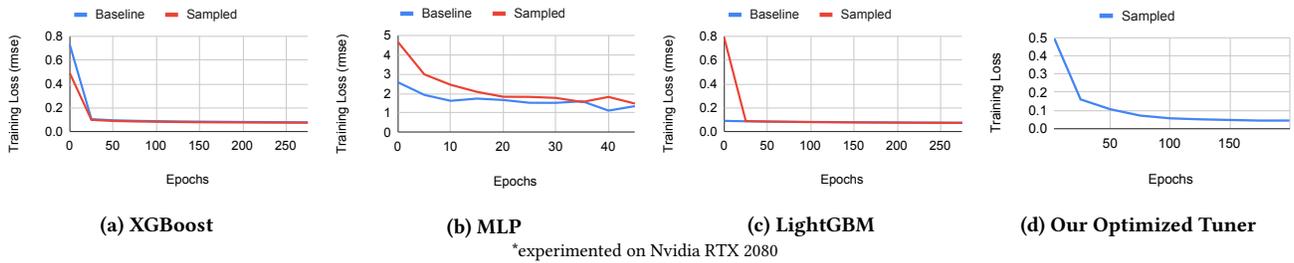

     \centering
     \captionsetup[subfigure]{justification=centering}
     
     \begin{subfigure}[]{0.24\textwidth}
         \centering
         \includegraphics[width=\textwidth]{assets/xgb.png}
        \caption{XGBoost}
        \label{fig:xgboost}
     \end{subfigure}
     ~
     \begin{subfigure}[]{0.24\textwidth}
        \centering
        \includegraphics[width=\textwidth]{assets/mlp.png}
        \caption{MLP}
        \label{fig:mlp}
     \end{subfigure}
     ~
     \begin{subfigure}[]{0.24\textwidth}
         \centering
        \includegraphics[width=\textwidth]{assets/lgbm.png}
        \caption{LightGBM}
        \label{fig:lgdm}
     \end{subfigure}
    ~
    \begin{subfigure}[]{0.24\textwidth}
        \centering
        \includegraphics[width=\textwidth]{assets/our_tuner.png}
        \caption{Our Optimized Tuner}
        \label{fig:our_tuner}
     \end{subfigure}
     \footnotesize{*experimented on Nvidia RTX 2080}
     \vspace{-0.2cm}
     \caption{Comparing Training Convergence of Tuners}
     \label{fig:convergence}
\end{figure*}

We have also compared the tuners for the epochs required to converge on the baseline and sampled data. As per the design of TVM's auto-scheduler, if they are executed for a large number of trials, evidently, the tuners, like XGB and MLP, will converge. To have a fair comparison, we have compared them by epochs. As shown in Figure 2, there is not much difference for XGB, MLP, and LGBM tuners on either dataset. On the other hand, our attention-inspired tuner performed much better by converging in a similar number of epochs but achieving twice the better error loss. The rmse for our optimized tuner is 0.04 after 200 epochs compared to 0.08 and 0.09 of XGB and LGBM, respectively. However, we are addressing an offline training overhead involved here as part of our next steps. Additionally, this is our first step, and we are also researching the instability of the tuners.

\section{Conclusion and Future Directions}
%% The conclusion is  not in sync with the abstract (  Just a comment)
In this research, we have demonstrated the effectiveness of the neural network and hardware parameters-aware sampling in automating tensor program generation for search-based tensor compilers. We showed the impact of various split strategies on the end-to-end optimization duration and early convergence. Mapping tensor operators to specific hardware may be crucial in a heterogeneous environment. Here, we have integrated hardware features into the evolutionary search procedure for efficient tensor program generation. We concluded that a heterogeneous features-aware training strategy could reduce training overhead regarding dataset requirements and yield effective transfer learning with fewer online measurements. 
After presenting our preliminary results, we intend to research selective feature training during transfer learning. Our future work includes improving the efficiency of cross-device and inter-subgraph learning with an evaluation of a scientific application. 

\begin{acks}
\small
%MLHPC project \\
{\footnotesize
This research was supported in part by the Exascale Computing Project (17-SC-20-SC), a collaborative effort of the U.S. Department of Energy Office of Science and the National Nuclear Security Administration. This material is
also based upon work supported by the National Science Foundation under
grant no. CCF-2113996. 
This research used resources of the Argonne Leadership Computing Facility (ALCF), which is a DOE Office of Science User Facility supported under Contract DE-AC02-06CH11357}
\end{acks}

\bibliographystyle{ACM-Reference-Format}
%\pagebreak 
\balance
\bibliography{ref}

\end{document}